\UseRawInputEncoding
\documentclass[10pt,final,doubleside]{IEEEtran}
	
	\usepackage{pifont,bm,multicol,amsfonts,amsmath,color,amssymb,graphicx, epsfig,cite,psfrag,subfigure,algorithm,enumerate,stfloats,algorithmic,epstopdf,balance,multirow,multicol}
\usepackage{booktabs}

\begin{document}

\title{Deep Learning-based Time-varying Channel Estimation for RIS Assisted Communication}

\author{Meng Xu, Shun Zhang, \emph{Senior Member, IEEE}, Jianpeng Ma, \emph{Member, IEEE}, Octavia A. Dobre, \emph{Fellow, IEEE}
\thanks{M. Xu, S. Zhang and J. Ma are with the State Key Laboratory of Integrated Services Networks, Xidian University, Xi¡¯an 710071, P. R. China (e-mail: mxu$\_$20@stu.xidian.edu.cn, zhangshunsdu@xidian.edu.cn, jpmaxdu@gmail.com).}

    \thanks{O. A. Dobre is with Faculty of Engineering and Applied Science, Memorial University, St. John's NL A1B 3X5, Canada (e-mail: odobre@mun.ca).}

}
\maketitle

\vspace{-15mm}

\begin{abstract}

Reconfigurable intelligent surface (RIS) is considered as a revolutionary technology for future wireless communication networks. In this letter, we consider the acquisition of
the time-varying cascaded channels, which is a challenging task due to the
massive number of passive RIS elements and the small channel coherence time. To reduce the pilot
overhead, a deep learning-based
channel extrapolation is implemented  over both antenna and time domains.
We divide the neural network
into two parts, i.e., the time-domain  and the antenna-domain  extrapolation networks, where
the neural ordinary differential equations (ODE) are utilized. In the former, ODE
accurately describes the dynamics of the RIS channels and improves the recurrent neural
network's performance of time series reconstruction. In the latter, ODE is resorted to
modify the relations among different  data layers in a feedforward neural network.
We cascade the two networks and
jointly train them. Simulation results show that the proposed
scheme can effectively extrapolate the cascaded RIS channels in high mobility scenario.

\end{abstract}

\maketitle
\thispagestyle{empty}
\vspace{-1mm}

\section{Introduction}
Massive and diversified communication services
put forward higher requirements, such as low energy cost and full coverage, to the upcoming 6G communication system.
As a promising new technology, reconfigurable intelligent surface (RIS) has attracted more and more attentions.
With the artificial electromagnetic structure, RIS can actively customize the wireless propagation link. Specially,
by applying control signals to the tunable elements on the electromagnetic units, the electromagnetic properties of these units can be controlled dynamically \cite{RIS}. Moreover, RIS can work in the passive model, which can greatly decrease the system's power consumption.
Recent studies demonstrate that RIS can improve the quality of the received signal, expand the coverage range and enhance the capacity of the wireless network \cite{RIS2} -\cite{RIS3}.

In order to fully embrace the above advantages of RIS, accurate  channel state information (CSI) should be acquired.
In \cite{channel_estimation1}, Liu \emph{et al.} proposed a message passing-based algorithm to factorize the cascaded channels.
The authors in \cite{channel_estimation3} adopted a two-stage channel estimation scheme by using atomic norm minimization to sequentially estimate the channel parameters.
In \cite{channel_estimation2}, Kim \emph{et al.} proposed a single-path approximated channel and selective emphasis on rank-one matrices to enable practical IRS-empowered SU-MIMO systems with low training overhead.
As mentioned in \cite{channel_estimation1} -\cite{channel_estimation2}, the main challenge for the channel estimation over RIS networks comes from
the large number of passive reflection elements at the RIS node. In order to decrease the channel estimation overhead, more and more researchers
try to  exploit the non-linear mapping between channels either at partial or all RIS elements and to implement effective channel compression over the antenna space.
Due to the universal approximation ability of the neural networks, deep learning (DL)-based channel extrapolation frameworks have been designed to infer the full channels from the partial ones over the antenna domain.
The authors in \cite{extrapolation} constructed a convolutional neural network (CNN)-based
framework to complete the channel extrapolation over the antenna domain.
Gao \emph{et al.} developed a three-stage training strategy and utilize both fully connected network and CNN to realize the extrapolation task in \cite{DL1}.

However, in practice, the users can move and
the channels between RIS and users would vary in time. As is well known, the
higher the mobility speed is, the lower the channel coherence
time is. Then, it would be more challenging to acquire a
large number of unknown RIS channels within a limited channel
coherence time. In this scenario, to ensure the system's spectrum efficiency, we will utilize as few pilot symbols as possible to achieve partial channel information within a given time interval. Hence, the idea of the channel extrapolation over the antenna domain could be applied for time-domain as well.


Thus, in this paper, we consider the time-varying cascaded channel estimation over RIS-assisted communication. We resort to DL and
utilize the idea of channel extrapolation in both antenna and time domains.
Correspondingly, the entire neural network
can be divided into two parts, i.e., the time-domain  and the antenna-domain  extrapolation networks.
Specially, in the former, we merge the recurrent neural
network (RNN) with the neural ordinary differential equations (ODE), which can accurately describe the dynamics of the RIS channels. In the latter, we utilize another function of the ODE, i.e., modifying the structure of neural networks, and design an enhanced feedforward neural network (FNN)
to achieve better extrapolation performance.
Then, we cascade the two networks and design a  training scheme to jointly optimize them.


\vspace{-2mm}
\section{System And Channel Model}
Let us consider a scenario where a base station (BS) communicates with a single antenna user equipment (UE) via RIS.
The BS's $M$ antennas are in the form of uniform linear array (ULA), and
RIS is equipped with $N$ reflective elements in the form of a uniform planar array (UPA), including $N_v$ the vertical direction and $N_h$ in the horizontal direction.
The links from the BS to UE include the direct link and the cascaded one via RIS.
The cascaded link consists of the channel from BS to RIS and that from RIS to UE.
Without loss of generality, as BS and RIS are placed at fixed positions with limited local scattering, the channel between them can be considered to be a light-of-sight (LoS) and time-invariant link under a long time interval
and can be written as
\begin{align}
\mathbf{H} = \sqrt{MN}\alpha \mathbf{a}_A(\psi) \mathbf{a}_R^{\mathrm{H}}(\phi_{h},\varphi_{h})\in \mathbb{C}^{M\times N}, \label{1}
\end{align}
where $\alpha$ is the complex channel gain.
$\mathbf{a}_R(\phi,\varphi)$ and $\mathbf{a}_A^{\mathrm{H}}(\psi)$ denote the steering vectors of the RIS and BS, respectively, with $\phi$ and $\varphi$ as the azimuth angle and elevation angle of RIS, and $\psi_{h}$ as the angle of departure (AoD).
$
\mathbf{a}_R(\phi_{h},\varphi_{h}) = \mathbf{a}_{el}(\varphi_{h})\otimes \mathbf{a}_{az}(\phi_{h},\varphi_{h})\in \mathcal{C}^{N\times 1},
$
$
\mathbf{a}_A(\psi_{h}) = [1, e^{j\frac{2\pi}{\lambda}d\sin\psi_{h}}, \dots, e^{j\frac{2\pi}{\lambda}d(M-1)\sin\psi_{h}}]^\mathrm{T},
$
where the $N_v\times 1$ vector $\mathbf{a}_{el}(\varphi_{h}) = [1, e^{j2\pi\frac{d}{\lambda}\cos\varphi_{h}}, \dots,  e^{j2\pi\frac{d}{\lambda}(N_v-1)\cos\varphi_{h}}]^\mathrm{T}$ and the $N_h\times 1$ vector $\mathbf{a}_{az}(\phi_{h},\varphi_{h}) = [1, e^{j2\pi\frac{d}{\lambda}\sin\phi_{h}\cos\varphi_{h,i}}, \dots, e^{j2\pi\frac{d}{\lambda}(N_h-1)
\sin\phi_{h}\cos\varphi_{h}}]^\mathrm{T}$.
$\lambda$ is the carrier wavelength and $d$ denotes the antenna spacing. Furthermore, $\otimes$ is the Kronecker product operator and $[\cdot]^\mathrm{T}$ represents the transpose.

Due to the mobility of UE, the channel between RIS and UE experiences time-selective fading.
Without loss of generality, we assume that the channel is quasi-static during a time block of $L_c$ channel uses and changes
from block to block.
Then, the value at the $n$-th time block is

\vspace{-3mm}
\begin{small}\begin{align}
\mathbf{g}_n = \sqrt{\frac{N}{L_g}}\sum_{i=1}^{L_g}\beta_i e^{j2\pi (n \frac{vf}{c}\cos\theta_i L_c T_s - f\tau_i)} \mathbf{a}_R(\phi_{g,i},\varphi_{g,i}) \in \mathbb{C}^{N\times 1}, \label{g}
\end{align}\end{small}where $\beta_i$ is the complex channel gain along the $i$-th path and $L_g$ is the number of scattering paths.
$v, f, c$ and $T_s$ separately represent the moving speed of UE, the carrier frequency, the speed of light, and the system sampling period.
$\tau_i$ and $\theta_i$ denote the time delay
and the angle between the incident direction of the electromagnetic wave and the movement direction of UE for the $i$-th path.

Thus, at time $n'$ of the $n$-th time block,
the received signal at BS side is expressed as
\begin{align}
\mathbf{y}(n') =\mathbf{H} \mathbf{\Phi}(n^\prime) \mathbf{g}(n) s(n') + \mathbf{v}(n'),
\end{align}
where the diagonal matrix $\mathbf{\Phi}(n^\prime)$ denotes the amplitude and phase control information at  RIS, i.e.,
$\mathbf{\Phi}(n^\prime) =\text{diag}\{\beta_1(n^\prime) e^{j\vartheta_{1}(n^\prime)}, \beta_2(n^\prime) e^{j\vartheta_{2}(n^\prime)}, \cdots, \beta_N(n^\prime) e^{j\vartheta_{N}(n^\prime)}\} \in \mathbb{C}^{N\times N}$, $s(n')$ is the user's transmitting data, and $\mathbf{v}(n')$ denotes the additive white Gaussian noise
with zero-mean and variance $\sigma_n^2$.
In order to represent the cascaded channel more clearly, the received signal can be written as
\begin{align}
\mathbf{y}(n') = \underbrace{\mathbf{H} \text{diag}\{\mathbf{g}(n)\}}_{\mathbf{C}(n)} \boldsymbol{\rho}(n^\prime) s(n') + \mathbf{v}(n'),
\label{eq:y_n}
\end{align}where $\mathbf C(n)$ is the cascaded channel of size $M\times N$, and the $N\times 1$  vector $\boldsymbol\rho(n^\prime)$
is formed by the diagonal elements of $\boldsymbol{\Phi}(n^\prime)$.
We can obtain the vector $\mathbf c(n)\in \mathbb{C}^{MN\times 1}$ by vectorizing $\mathbf C(n)$.
Notice that $n^\prime$ in $\mathbf y(n^\prime)$, $\mathbf v(n^\prime)$, $\boldsymbol\Phi(n^\prime)$, $\boldsymbol\rho(n^\prime)$
and $s(n^\prime)$ denotes the instant time index, while $n$ in $
\mathbf C(n)$, $\mathbf G(n)$ represents the block index.


\section{DL-Based Channel Estimation Network}
\vspace{-1mm}
\subsection{Proposed Problems}

\begin{figure}[!t]
	\centering
	\includegraphics[width=3.5in]{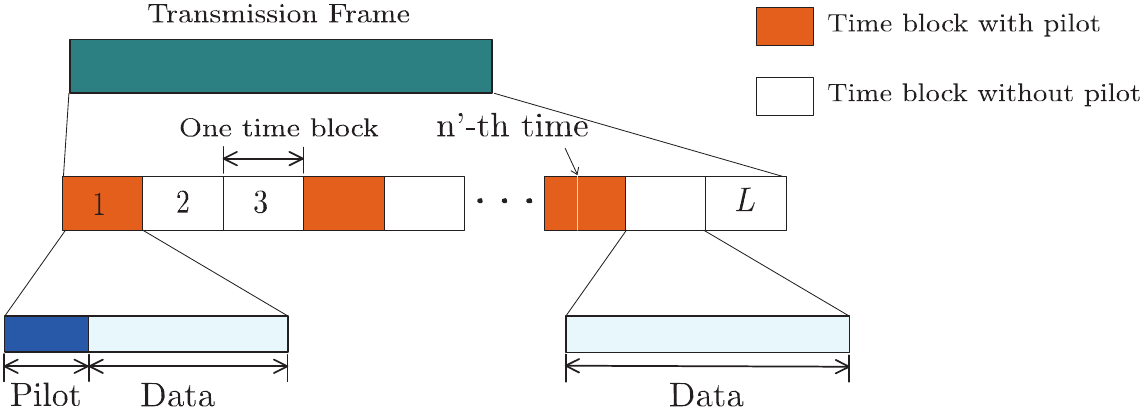}
	\caption{{The structure of the transmission frame.}}
	\label{Pilot pattern}
\end{figure}

We assume that each uplink frame from UE to BS  contains $L$  time blocks as shown in Fig. \ref{Pilot pattern}.
Before proceeding, let us define $\mathcal L=\{1,2,\ldots,L\}$ and
$\mathcal A=\{1,2,\ldots, N\}$.
To implement data detection, BS should recover
$L$ matrices of size $M\times N$, i.e.,
$\{\mathbf C(n)|n\in \mathcal L\}$.
In the RIS assisted communication with time-selective fading, the estimation of $\{\mathbf C(n)|n\in \mathcal L\}$ faces the following problem.
At each time block, the size of the cascaded channel $\mathbf C(n)$ is proportional to the number of RIS elements, i.e., $N$.
To acquire $\mathbf C(n)$, the length of the required pilot sequence is also proportional to $N$.
In order to reduce this length, fewer RIS elements could be selected to implement the channel compression by controlling each RIS element's on/off state during each pilot block.
Without loss of generality, we assume that the  RIS elements during these pilot blocks have the same on/off pattern, and
the  indexes  of all selected RIS elements are collected into the set $\mathcal A^p$, where $|\mathcal A^p| = N_s \ll N$.
Correspondingly, the amplitude control information of the selected RIS elements in $\mathcal A^p$ is 1, while for the others is 0, i.e., $\beta_i(n^\prime)=1, i\in\mathcal A^p$ and $\beta_j(n^\prime)=0,~j\notin\mathcal A^p$.
Then, within each pilot block, our objective channel becomes matric of size $M\times N_s$, i.e.,
$\{\mathbf C_{:,\mathcal A^p}(n)|n\in \mathcal L\}$.
Moreover,  to ensure the system's spectrum efficiency,  the number of pilot symbols should not be too large, which means that it is not necessary to insert pilot sequences in each time block.
Define the set of time blocks for insertion of pilots as $\mathcal L^p$.
Then, the $n$-th pilot block can be utilized to achieve the information about $\{\mathbf C_{:,\mathcal A^p}(n)|n\in \mathcal L^p\}$.

Due to the fixed RIS structure and channels' time correlation, the following mapping between
$\{\mathbf C_{:,\mathcal A^p}(n)|n\in \mathcal L^p\}$ and $\{\mathbf C(n)|n\in \mathcal L\}$ exists
\begin{align}
\bm{\Phi}_H^i:\{\mathbf C_{:,\mathcal A^p}(n)|n\in \mathcal L^p\}\rightarrow  \{\mathbf C(n)|n\in\mathcal L\}.\label{eqn:inter_map}
\end{align}

Since the spatial and time correlation of $\mathbf c(n)$ are uncoupled, the above mapping can be achieved through
two sequential operations

\vspace{-3mm}
\begin{small}\begin{align}
\bm{\Phi}_H^t&:\{\mathbf C_{:,\mathcal A^p}(n)|n\in \mathcal L^p\}\rightarrow  \{\mathbf C_{:,\mathcal A^p}(n)|n\in\mathcal L\}, \\
\bm{\Phi}_H^a&:\{\mathbf C_{:,\mathcal A^p}(n)|n\in\mathcal L\}\rightarrow  \{\mathbf C(n)|n\in\mathcal L\}, \label{task}
\end{align}\end{small}where the former denotes the channel interpolation along the time-dimension, while the latter is the channel extrapolation over the antenna-domain.

\subsection{Initial Cascaded Channel Estimation}

In this part, we will resort to a simple linear estimator to achieve coarse information about $\{\mathbf C_{:,\mathcal A^p}(n)|n\in \mathcal L^p\}$.
Without loss of generality, we assume that the pilot symbol  from  the user at the $i$-th time block  is $\sqrt{\frac{P}{N^p}}$, where $i\in\mathcal L^p$, $N^p$ is the length of pilot sequence and $P$ is the pilot power during this time block.
Let us collect $N^p$ observation vectors of size $M\times1$ during the $i$-th time block inserted into $M\times N^p$ matrix $\mathbf Y_i^p=[\mathbf y(n_i^s), \mathbf y(n_i^s+1),\ldots,\mathbf y(n_i^s+N^p-1)]$, where $n_i^s$ denotes the left border of the pilot sequence in the $i$-th time block, $i\in\mathcal L^p$.
Their corresponding $N^p$ control vectors of size $|\mathcal A^p|\times 1$ are placed into the
$|\mathcal A|\times N^p$ matrix $\boldsymbol\Gamma_i=[\boldsymbol\rho_{\mathcal A^p}(n_i^s), \boldsymbol\rho_{\mathcal A^p}(n_i^s+1),\ldots,\boldsymbol\rho_{\mathcal A^p}(n_i^s+N^p-1)]$. Then,
with (\ref{eq:y_n}), $\mathbf Y_i^p$ can be written as
\begin{align}
\mathbf Y_i^p=\sqrt{\frac{P}{N^p}} \mathbf C_{:,\mathcal A^p}(n_i^{s})\boldsymbol\Gamma_i+\mathbf V_i^p,
\end{align}
where the $M\times N^p$ matrix $\mathbf V_i^p=[\mathbf v(n_i^s), \mathbf v(n_i^s+1),\ldots,\mathbf v(n_i^s+N^p-1)]$. Let $\boldsymbol\Gamma_i\boldsymbol\Gamma^H=\mathbf I_{|\mathcal A|}$ and $N^p\ge|\mathcal A|$. Then, we can achieve the coarse estimation of
$\mathbf{C}_{:,\mathcal A}(i)$ as
\begin{align}
\mathbf{\overline C}_{:,\mathcal A^p}(i)=\sqrt{\frac{N^P}{P}}\mathbf Y_i^p\boldsymbol\Gamma_i^H,\kern 10pt i\in\mathcal L^p,
\end{align}
where $\mathbf{\overline C}_{:,\mathcal A^p}(i)$ denotes the initial information of $\mathbf{C}_{:,\mathcal A^p}(i)$.
Similar to the relation between $\mathbf C(i)$ and $\mathbf c(i)$, we define the corresponding vector version of  $\mathbf{\overline C}_{:,\mathcal A^p}(i)$ as $\mathbf{\bar c}_{\mathcal A^p}(i)$.
Then, we can achieve the coarse information about $\{\mathbf C_{:,\mathcal A^p}(i)|i \in \mathcal L^p\}$.

\vspace{-3mm}
\subsection{DL-based Spatial Extrapolation and Temporal Interpolation for RIS Channels}

We first consider the mapping $\bm{\Phi}_H^t$ in (\ref{task}).
Since RNN can effectively capture the time sequence's dynamical characteristics, we adopt it here.
Before proceeding, let us define the raw input of RNN as $\mathbf{x}(n)$. Moreover, when $n \in \mathcal L^p$, $\mathbf{x}(n) = \mathbf{\bar c}_{\mathcal A^p}(n)$, and if $n \in \mathcal L - \mathcal L^p$, $\mathbf{x}(n) = \bold{0}$.

For a given time sequence set $\{\mathbf  x(n)| n\in\mathcal L\}$, RNN would extract
the hidden dynamical state set  $\{\mathbf u(n)|n\in\mathcal L\}$. Hence, RNN-based channel interpolation contains
two function blocks.
The first one updates the hidden state $\mathbf{u}(n)$ utilizing RNN with parameters $\boldsymbol\omega_R$, which is denoted as ``$\text{RNNCell}_{\boldsymbol\omega_R}$''.
Here, $\mathbf{x}(n)$ is separately put into $\text{RNNCell}_{\boldsymbol\omega_R}$ in the chronological order by taking $n$ from $1$ to $L$. At each $n$,
$\text{RNNCell}_{\boldsymbol\omega_R}$ deals with the current raw input $\mathbf x(n)$ and the previous hidden state $\mathbf u(n-1)$, and outputs
the current state $\mathbf{u}(n)$.
Correspondingly, the second block needs to infer $\mathbf c_{\mathcal A^p}(n)$ from the hidden state $\mathbf u(n)$, where
a decoding network with parameters $\boldsymbol\omega_D$ ($\text{DecNet}_{\boldsymbol\omega_D}$)  is utilized. In fact, the output of $\text{DecNet}_{\boldsymbol\omega_D}$ at time $n$ is the estimation of $\mathbf c_{\mathcal A^p}(n)$, i.e., $\widehat{\mathbf c}_{\mathcal A^p}(n)$.
Then, RNN-based channel interpolation can be formulated as
\begin{align}
\begin{cases}
\mathbf{u}(n) = \text{RNNCell}_{\boldsymbol\omega_R}(\mathbf{u}(n-1), \mathbf{x}(n)), \label{RNNCell} \\
\widehat{\mathbf{c}}_{\mathcal A^p}(n) = \text{DecNet}_{\boldsymbol\omega_D}(\mathbf{u}(n)),\kern 20pt n\in\mathcal L.
\end{cases}
\end{align}

As we can observe from (\ref{RNNCell}),
$\mathbf u(n)$ remains the same within the $n$-th and $(n-1)$-th time blocks, which does not fit the $\mathbf c_{\mathcal A^p}(n)$'s dynamical characteristics, especially when the irregular pilot blocks are inserted.
To deal with this problem, we resort to the ODE
and model the dynamical hidden state $\mathbf u(n)$. Theoretically, ODE can be seen as a continuous-time model and
can be formulated as
\begin{align}
\frac{d \xi(t)}{dt} = f(\xi(t),t), \label{ODE}
\end{align}
where $\xi(t)$ is the continuous dynamical state and $f(\cdot)$ specifies the dynamics of $\xi(t)$.
In neural networks, $f(\cdot)$ can be approximated by a simple network with parameters $\boldsymbol \omega_f$.
Then, $f(\cdot)$ can be written as $f_{\boldsymbol \omega_f}(\cdot)$.
With a given $f_{\boldsymbol \omega_f}(\cdot)$ and the initial value $\xi(0)$,
the numerical ODE solver,  i.e., ``ODESolver,'' can be utilized to evaluate the specific values of $\xi(t)$ at any desired time set $\{t_0,t_1,\ldots,t_{T-1}\}$ as

\vspace{-3mm}
\begin{small}
\begin{align}
\{\xi(t_0), \xi(t_1), \cdots, \xi(t_T)\} = \text{ODESolver}(f_{\boldsymbol \omega_f}, \xi(0), \{t_0, \cdots, t_T\}). \label{ODESolver}
\end{align}
\end{small}

\begin{figure}[!t]
	\centering
	\includegraphics[width=3.5in]{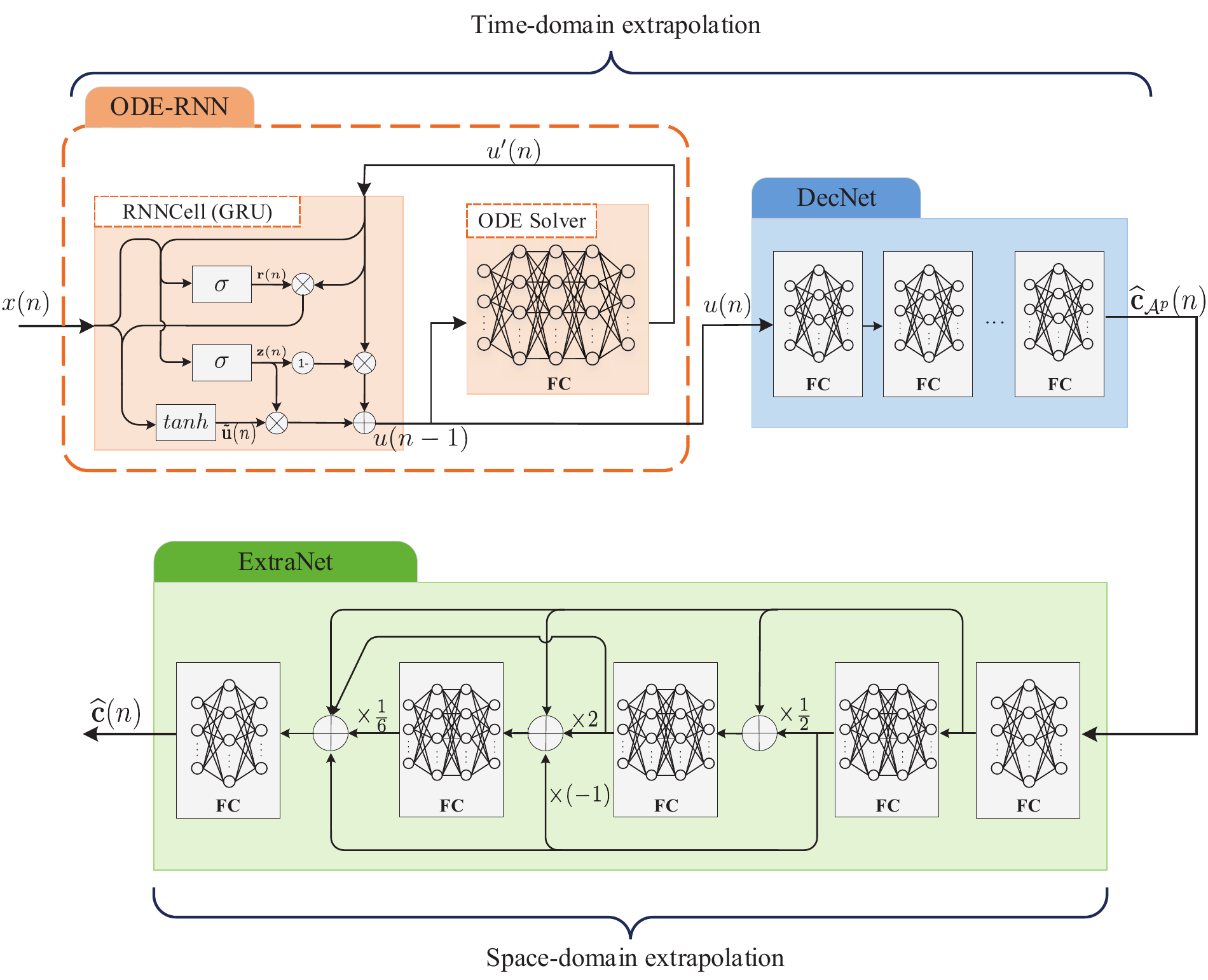}
	\caption{{The architecture of the proposed neural network for channel estimation.}}
	\label{network}
\end{figure}
Plugging ODESolver into (\ref{RNNCell}),  we formulate the  ODE-RNN network structure, which can implement $\bm{\Phi}_H^t$ as
\begin{align}
\begin{cases}
\mathbf{u}'(n) = \text{ODESolver}(f_{\boldsymbol \omega_f}, \mathbf{u}(n-1), (t_n, t_{n-1})), \\
\mathbf{u}(n) = \text{RNNCell}_{\boldsymbol\omega_R}(\mathbf{u}'(n), \mathbf{x}(n)), \\
\widehat{\mathbf{c}}_{\mathcal A^p}(n) = \text{DecNet}_{\boldsymbol\omega_D}(\mathbf{u}(n)),
\end{cases}
\end{align}
where $\mathbf{u}'(n)$ is the middle hidden state output by ODESolver.

In particular, we adopt a fully connected neural network as the network of ODESolver and DecNet, and adopt the Gated Recurrent Unit (GRU) as hidden state update formula for the RNNCell function, which is defined as follows:

\vspace{-3mm}
\begin{small}\begin{flalign}
&\mathbf{r}(n) = \sigma(W_r ([\mathbf{u}^{\prime}(n), \mathbf{x}(n)])), \kern 58pt \text{reset gate}\\
&\mathbf{z}(n) = \sigma(W_z ([\mathbf{u}^\prime(n), \mathbf{x}(n)])), \kern 58pt \text{update gate}\\
&\tilde{\mathbf{u}}(n) = \tanh(W_{\tilde{u}} ( [\mathbf{r}(n) \odot \mathbf{u}^\prime(n),\mathbf{x}(n)])),  \kern 15pt\text{new memory}\\
&\mathbf{u}(n) = (1 - \mathbf{z}(n)) \odot \mathbf{u}^\prime (n)+ \mathbf{z}(n) \odot \tilde{\mathbf{u}}(n), \kern 7pt \text{hidden state}
\end{flalign}\end{small}where $\sigma$ is the sigmoid function with the form of $\sigma(x) = \frac{1}{1 + e^{-x}}$,
and $W_r(\cdot)$, $W_z(\cdot)$ and $W_{\tilde{u}}(\cdot)$ are the network of reset gate, update gate, and new state update with different parameters, respectively.
Moreover,
$\mathbf{r}(n)$, $\mathbf{z}(n)$ and $\tilde{\mathbf{u}}(n)$ are outputs of corresponding steps, and
$[a, b]$ denotes the concatenation operation of $a$ and $b$.

Then, we consider the mapping $\bm{\Phi}_H^a$ in (\ref{task}), which corresponds to the
channel extrapolation process over the antenna-domain.
Similar to the super-resolution in the field of image processing,
we can use the neural network to approximate this mapping from $\widehat{\mathbf{c}}_{\mathcal A^p}(n)$ to $\widehat{\mathbf{c}}(n)$.
In the last task, ODE is employed to describe a dynamical changing process.
In fact, according to the numerical solutions of ODE, we can
 modify the neural network structure and achieve better performance~\cite{neuralODE}.
Here, we adopt a numerical solution of ODE, i.e., Runge-Kutta method, to modify the structure of feedforward neural network as the spatial extrapolation network \cite{ODECNN}.
Then, a network with parameters $\boldsymbol\omega_E$ called $\text{ExtraNet}_{\boldsymbol \omega_E}$ can complete the following mapping as
\begin{align}
\widehat{\mathbf{c}}(n) = \text{ExtraNet}_{\boldsymbol \omega_E}(\widehat{\mathbf{c}}_{\mathcal A^p}(n)). \label{ExtraNet}
\end{align}

Correspondingly, the proposed network architecture is shown in Fig. \ref{network}.
Further, the detailed layer parameters settings we adopted in simulation are shown in TABLE \ref{layer parameters}.

\newcommand{\tabincell}[2]{\begin{tabular}{@{}#1@{}}#2\end{tabular}}
\begin{table}[t]
		\centering
		\renewcommand{\arraystretch}{1.2 }
		\caption{Layer Parameters for the Proposed Model.}
		\label{layer parameters}
        \scalebox{0.8}{
		\begin{tabular}{c c c c }
			\hline
			          &Layer                    &Output size                                                       &Activation\\
			\hline
            \hline
            \specialrule{0em}{2pt}{2pt}
			ODESolver &$3 \times $FC layer    &$M_b \times |\mathcal L| \times (MN \times 2 \times{r_a})$         &Tanh \\
            \specialrule{0em}{2pt}{2pt}
            \hline
			\tabincell{c}{RNNCell\\ ($W_r(\cdot)$, $W_z(\cdot)$, $W_{\tilde{u}}(\cdot)$)}
                      &$6 \times $FC layer    &$M_b \times |\mathcal L| \times (MN \times 2 \times{r_a})$         &Tanh \\
            \hline
            \specialrule{0em}{2pt}{2pt}
			DecNet    &$6 \times $FC layer    &$M_b \times |\mathcal L| \times (MN \times 2 \times{r_a})$         &Tanh \\
            \specialrule{0em}{2pt}{2pt}
            \hline
            \specialrule{0em}{2pt}{2pt}
			ExtraNet  &$8 \times $FC layer    &$M_b \times |\mathcal L| \times (MN \times 2)$                     &Tanh \\
            \specialrule{0em}{3pt}{2pt}			
            \hline
            \hline	
		\end{tabular}}	
	\end{table}


\vspace{-5mm}
\subsection{Learning Scheme}

{As shown in the previous sub-section,} the raw input of the neural
network is $\{\mathbf{x}(n) | n\in \mathcal L\}$, and the total target is $\{\mathbf{c}(n) | n\in \mathcal L\}$.
Moreover, as mentioned above, the network is divided into two parts: time-domain
and antenna-domain extrapolation networks.
Accordingly, in the network training stage, targets and loss functions should be set respectively for the two sub-networks to achieve the purpose of realizing different functions.
Similarly, the target of time-domain extrapolation network, i.e.,
the ideal output of $\text{DecNet}_{\boldsymbol\omega_D}$, is
$\{\mathbf{c}_{\mathcal A^p}(n) | n\in \mathcal L\}$.

Let us define $\mathcal T$ as the network training dataset, where $|\mathcal T|=N_{tr}$ is
the number of training sample. One sample in $\mathcal T$ contains three matrix sequences that are denoted as $(\{\mathbf{x}(n) | n\in \mathcal L\},\{\mathbf{c}_{\mathcal A^p}(n) | n\in \mathcal L\},\{\mathbf{c}(n) | n\in \mathcal L\})$, where $\{\mathbf{c}_{\mathcal A^p}(n) | n\in \mathcal L\}$ and
$\{\mathbf{c}(n) | n\in \mathcal L\})$ are the labels of time-domain and antenna-domain extrapolation networks, respectively.
Without loss of generality, we use the mean square error (MSE) of the channel estimation as the loss function, which can be separately written as
\begin{align}
\mathcal{L}_t =&\frac{1}{M_bMN_s|\mathcal L|}\sum_{i=1}^{M_b}
\sum_{n\in \mathcal L}\begin{Vmatrix}\mathbf c_{\mathcal A^p}(n) - \mathbf{\widehat c}_{\mathcal A^p}(n) \end{Vmatrix}_F^2, \\
\mathcal{L}_a =&\frac{1}{M_bMN|\mathcal L|}\sum_{i=1}^{M_b}\sum_{n\in \mathcal L}\begin{Vmatrix}\mathbf c(n) - \mathbf{\widehat c}(n) \end{Vmatrix}_F^2,
\end{align}
where $||\mathbf{A}||_F$ is the $\mathrm{Frobenius}$ norm of matrix $\mathbf{A}$ and $M_b$ denotes the batch size for training.
The total loss function is the weighted sum of $\mathcal L_t$ and $\mathcal L_a$, i.e., $\mathcal L_s = \mathcal L_t + \gamma\mathcal L_a$, where and $\gamma$ is the weighted coefficient.
Here, the adaptive moment estimation (Adam) \cite{ADAM} algorithm is adopted to achieve the best network parameters $\boldsymbol \omega = \{\boldsymbol \omega_f, \boldsymbol \omega_R, \boldsymbol \omega_D, \boldsymbol \omega_E\}$,
which is controlled by  the learning rate $\eta$.

\vspace{-3mm}
\section{Simulation Results}
In this section, we evaluate the performances of the proposed time interpolation and space extrapolation scheme through numerical simulation results.
We first describe the communication scenario and the adopted dataset, and then introduce the parameter setting of the proposed neural network.
Finally, the simulation results for performance evaluation are shown and explained.

We consider an environment with a BS, RIS, and UE.
A BS equipped with two antennas ($M = 2$) communicates with the single antenna UE through RIS, which has $64$ reflection elements ($N = 64$).
It is assumed that the positions of BS and RIS are fixed, while UE can move at a high speed which is set as $100$ km/h.
The generation of  the data is based on the DeepMIMO dataset \cite{DeepMIMO},
where the outdoor ray-tracing scenario 'O1' is adopted.
The parameters $\alpha, \psi, \phi_{h}, \varphi_{h}$ can be extracted from the 'O1' scenario to generate the time-invariant channel $\mathbf{H}$.
For generating the time-varying channel sample, we adopt the parameters $L_g, \beta_i, \phi_{g,i}, \varphi_{g,i}$ from DeepMIMO, and randomly select the angle between UE's movement and the direction of incident electromagnetic $\theta_i$ ($\theta_i \in [-20^\circ, 20^\circ]$). Then, we can generate the time-varying channel sample of each user according to (\ref{g}).
Furthermore, the carrier frequency of channel estimation is $28$ GHz, and the system bandwidth is set as $20$ MHz. The antenna spacing is $\frac{\lambda}{2}$ ($d = \frac{\lambda}{2}$) and the number of paths is set as $5$ ($L_g = 5$).
The time-domain sampling rate $r_t = \frac{|\mathcal L^p|}{|\mathcal L|}$ is set as a value in the set $\{0.3, 0.5, 1\}$, and the space-domain sampling rate $r_a = \frac{N_s}{N}$ is chosen in $\{\frac{1}{2}, \frac{1}{4}, \frac{1}{8}, \frac{1}{16}\}$.

The total number of cascaded channel samples is $20000$.
We employ $80\%$ of the data for training and the rest for testing.
When calculating the loss function, we set the weight coefficient $\gamma$ as $1$ to train the two networks jointly.
The total number of epochs is $1000$, and the batch size $M_b$ is $200$.
We set the initial learning rate $\eta$ as $0.005$, which decreases by $50\%$ for every $50$ epochs, and the lowest learning rate as $0.00005$.

\begin{figure}[!t]
	\centering
	\includegraphics[width=3.3in]{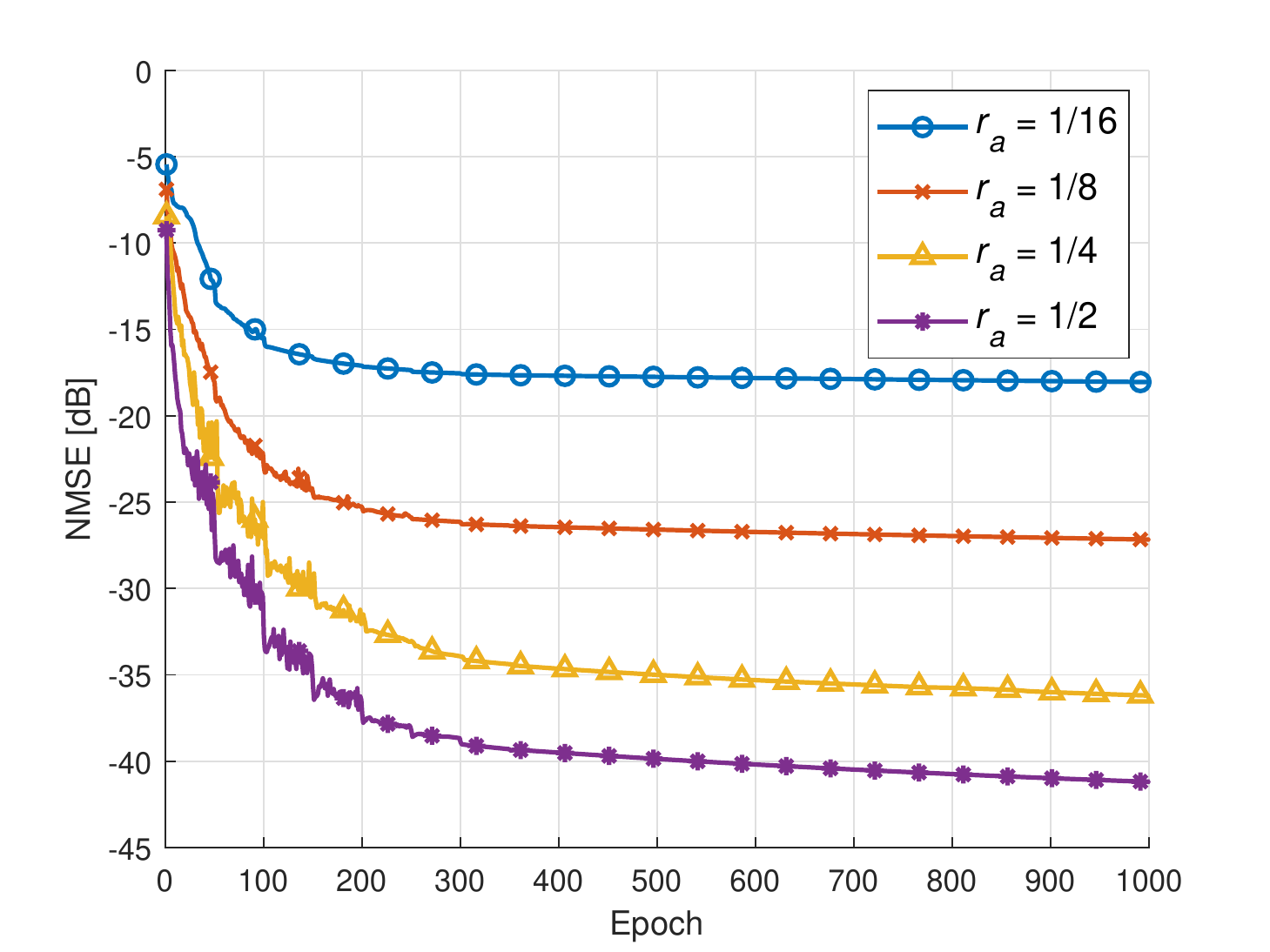}
	\caption{{NMSE of channel extrapolation vs. network training epoches for different $r_a$ values.}}
	\label{epoch_nmse}
\end{figure}

Fig. \ref{epoch_nmse} depicts the variation of normalized MSE (NMSE) versus epoch on the validation set at different spatial sampling rates. We adopt $\frac{1}{2}, \frac{1}{4}, \frac{1}{8}$ and $ \frac{1}{16}$ sampling rates and consider the case of no noise.
Obviously, it can be checked that the NMSE decreases with the epoch for all $r_a$. Further, it achieves the steady state within $1000$ epochs, which proves the robustness of the proposed scheme.
Moreover, with the increase of the number of RIS reflection elements, the performance of the proposed channel extrapolation scheme becomes better, converging to lower NMSE values.

\begin{figure}[!t]
	\centering
	\includegraphics[width=3.3in]{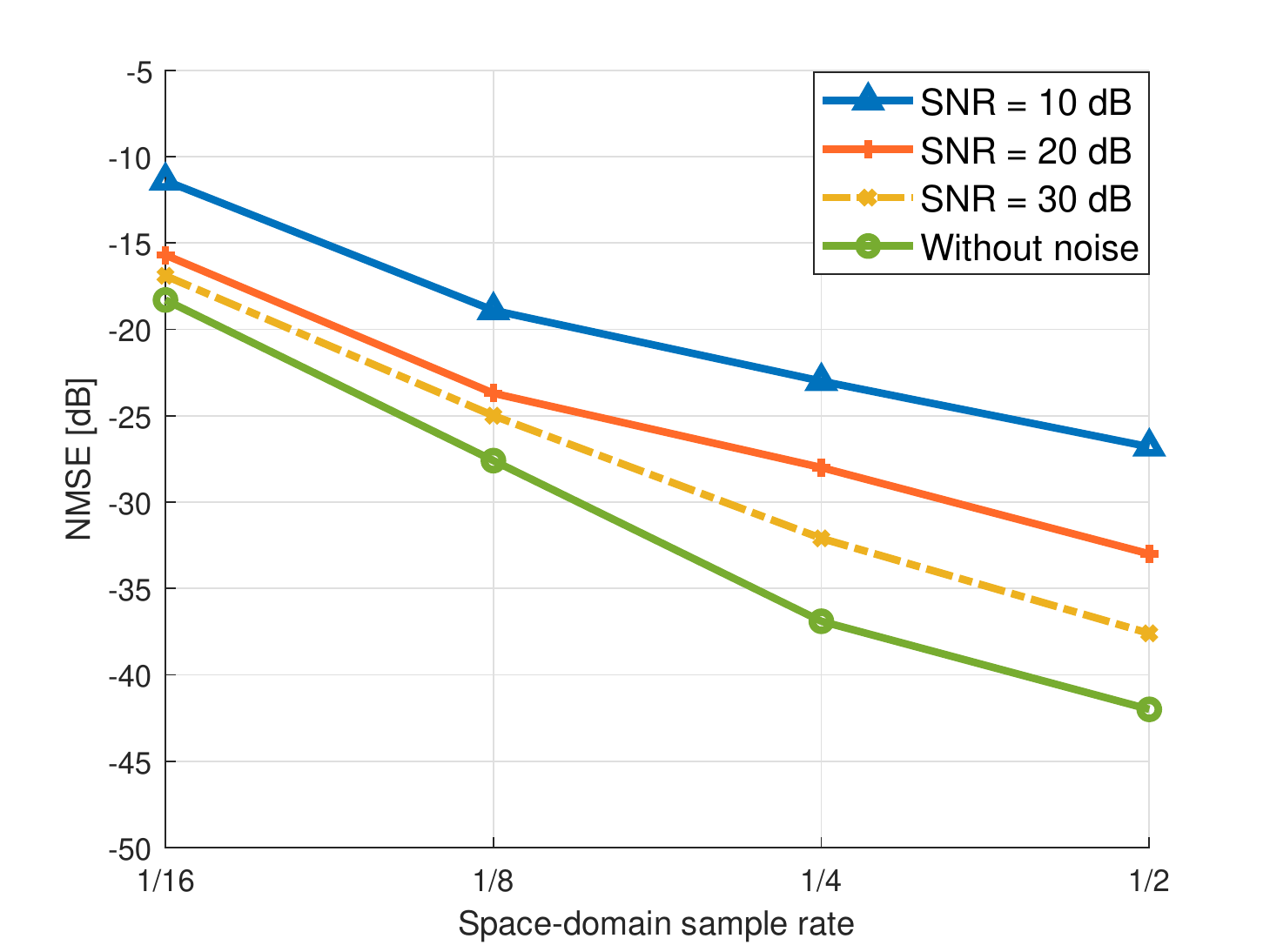}
	\caption{{NMSE of channel extrapolation vs. different antenna-domain sample rate for different SNR values.}}
	\label{r_noise}
\end{figure}

Fig. \ref{r_noise} shows the channel extrapolation performance versus the antenna-domain sampling rate under different SNRs.
The time-domain sampling rate $r_t$ is adopted as $0.3$.
As can be seen from the figure, with the decrease of SNR at the same sampling rate, the NMSE gradually increases and the channel extrapolation performance gradually deteriorates.
Furthermore, when noise is considered, the channel extrapolation performance becomes better with the increase of sampling rate, which is consistent with the results shown in Fig. \ref{epoch_nmse}.

\begin{figure}[!t]
	\centering
	\includegraphics[width=3.3in]{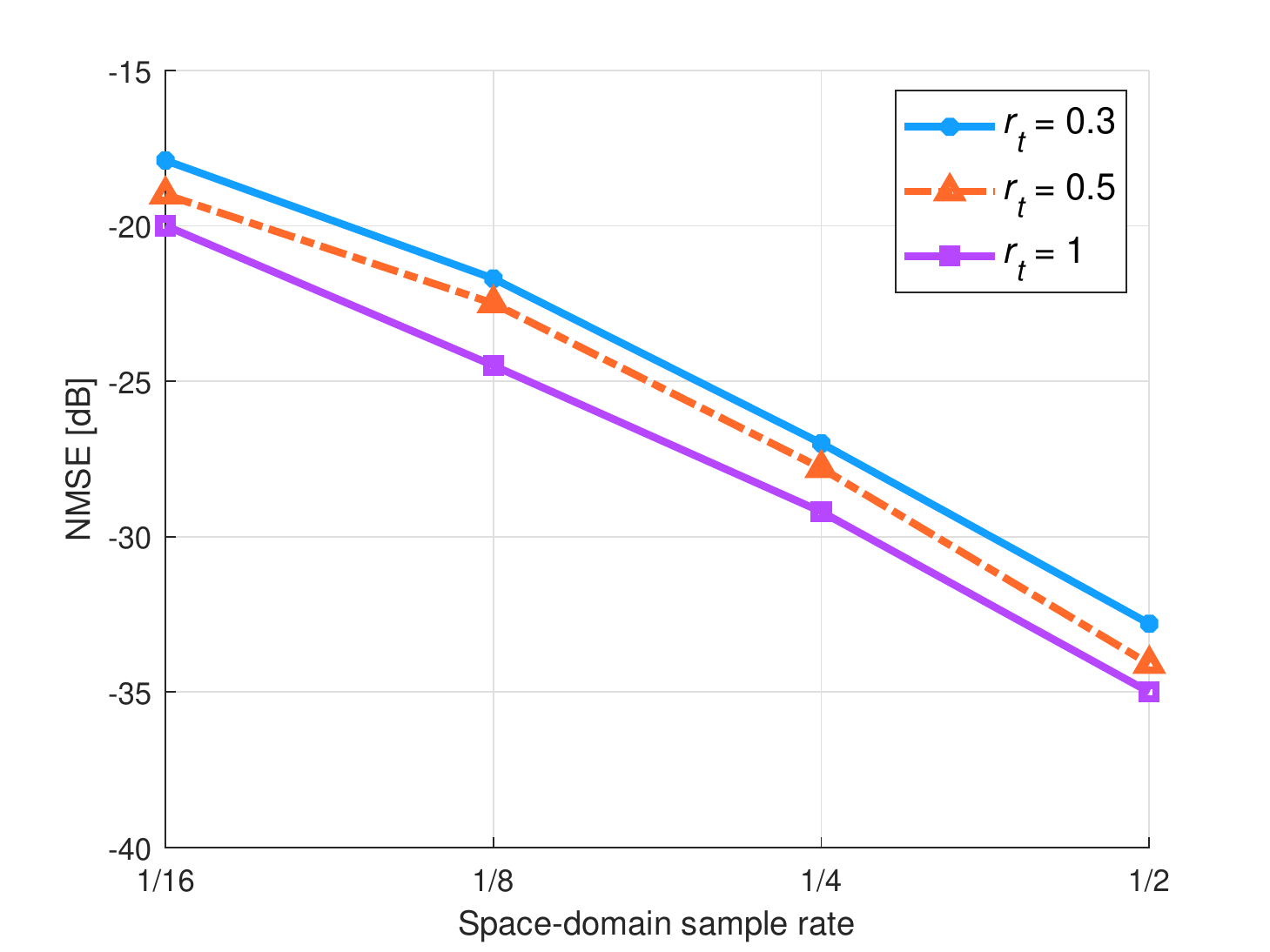}
	\caption{{NMSE of channel extrapolation vs. different space-domain sample rate for different $r_t$ values.}}
	\label{time_sample}
\end{figure}

Fig. \ref{time_sample} describes the extrapolation performance of the designed network at both spatial and temporal sampling rates.
We set SNR $20$ dB and investigate the network performance under $4$ different spatial sampling rates and $3$ different temporal spatial sampling rates, i.e., $r_a \in \{\frac{1}{2}, \frac{1}{4}, \frac{1}{8}, \frac{1}{16}\}$ and $r_t \in\{0.3, 0.5, 1\}$.
It is obvious that the NMSE decreases as the sampling rate increases, both for spatial and temporal sampling.
This can be explained as the smaller the number of sampling points is, the less information about the cascaded channels is available, which is not conducive for the designed network to extrapolate the channel.

\vspace{-3mm}
\section{Conclusion}

In this letter, we considered the time-varying channel acquisition problem in RIS scenario. To reduce the overhead of channel estimation, channel sub-sampling has been applied in both time and antenna domains, and a two-part cascaded neural network has been designed to accomplish channel interpolation in time-domain and channel extrapolation in antenna-domain through joint training.
Furthermore, ODE has been utilized to describe the dynamic process of temporal interpolation in the former part of the network and promote the network structure in the latter, i.e., the antenna extrapolation part.
Simulation results have illustrated that the cascaded channel extrapolation performance is satisfactory under the joint sampling of time and space domains, indicating that the proposed scheme is effective for time-varying channels and can also work well under the condition of noise, which proved its robustness.

\vspace{-5mm}

\linespread{1}

\end{document}